\documentclass{article}
\usepackage{multirow}
\usepackage{booktabs} \usepackage{subfigure}
\usepackage{algorithm}
\usepackage{algorithmicx}
\usepackage{algpseudocode}
\usepackage{amsfonts,amssymb}
\usepackage{spconf,amsmath,graphicx}
\usepackage{hyperref}

\hypersetup{
    colorlinks = true,
    linkcolor=blue,
    filecolor=blue,      
    urlcolor=blue,
    citecolor=cyan,
}

\def\x{{\mathbf x}}

\title{Simple Attention Module based Speaker Verification with Iterative noisy label detection}
\name{Xiaoyi Qin$^{1,2}$, Na Li$^{3}$, Chao Weng$^{3}$, Dan Su$^{3}$, Ming Li$^{1,2}$\thanks{Corresponding Author: Ming Li ming.li@whu.edu.cn. This research is funded in part by the National Natural Science Foundation of China (62171207), Tencent AI Lab Rhino-Bird Gift Fund, Fundamental Research Funds for the Central Universities (2042021kf0039), Key Research and Development Program of Jiangsu Province (BE2019054) and Science and Technology Program of Guangzhou City (201903010040,202007030011).}}
\address{$^{1}$School of Computer Science, Wuhan University, Wuhan, China \\
		$^{2}$Data Science Research Center, Duke Kunshan University, Kunshan, China\\
		$^{3}$Tencent AI Lab, Shenzhen, China}
\begin{document}
%
\maketitle
\begin{abstract}
Recently, the attention mechanism such as squeeze-and-excitation module (SE) and convolutional block attention module (CBAM) has achieved great success in deep learning-based speaker verification system. This paper introduces an alternative effective yet simple one, i.e., simple attention module (SimAM), for speaker verification. The SimAM module is a plug-and-play module without extra modal parameters. In addition, we propose a noisy label detection method to iteratively filter out the data samples with a noisy label from the training data, considering that a large-scale dataset labeled with human annotation or other automated processes may contain noisy labels. Data with the noisy label may over parameterize a deep neural network (DNN) and result in a performance drop due to the memorization effect of the DNN. Experiments are conducted on VoxCeleb dataset. The speaker verification model with SimAM achieves the 0.675\% equal error rate (EER) on VoxCeleb1 original test trials. Our proposed iterative noisy label detection method further reduces the EER to 0.643\%.
\end{abstract}

\label{sec:noisy label}

\begin{figure*}[htbp] \tiny
	\centering
	\subfigure [aWtugEAkhtM/00074]{
	\label{fig:spk1}
	\begin{minipage}[t]{0.2\linewidth}
	\centering
	\includegraphics[width=\linewidth]{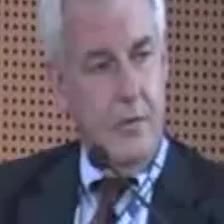}
	\end{minipage}
	}
	\subfigure[aWtugEAkhtM/00076]{
	\label{fig:spk2}
	\begin{minipage}[t]{0.2\linewidth}
	\centering
	\includegraphics[width=\linewidth]{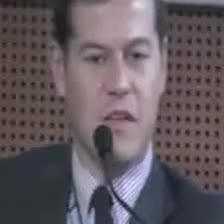}
	\end{minipage}
	}
	\subfigure[aWtugEAkhtM/00078]{
	\label{fig:spk3}
	\begin{minipage}[t]{0.2\linewidth}
	\centering
	\includegraphics[width=\linewidth]{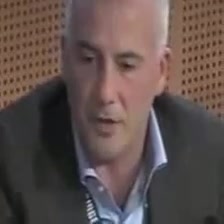}
	\end{minipage}
	}
	\subfigure[hqE1mX1V99k/000104]{
	\label{fig:spk4}
	\begin{minipage}[t]{0.2\linewidth}
	\centering
	\includegraphics[width=\linewidth]{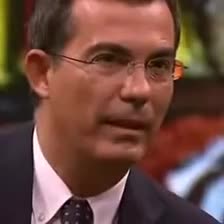}
	\end{minipage}
	}
	\caption{Visualization of noisy labeled faces. The four face images are all selected from speaker `id00244'. Figure (a), (b), (c) are from the `aWtugEAkhtM' segment and the (d) is from the `hqE1mX1V99k' segment. The face identity of (a) is dominant in the selected speaker and is considered as correct identity. Utterances from segment (b), (c) and (d) are with noisy labels.}
	\label{fig:error_spk}
\end{figure*}

\section{Introduction}
\label{sec:intro}
In the past few years, deep learning has significantly improved the performance of automatic speaker verification (ASV) systems. Neural network structures such as time-delay neural network (TDNN) \cite{xvector,ftdnn}, residual convolutional neural network (ResNet) \cite{resnet} and Res2Net \cite{res2net} have been explored and successfully applied to the ASV task. In addition to the improvement of the network structures, the availability of large-scale datasets as well as the carefully designed data augmentation strategies also improve the robustness of the ASV systems in many challenging scenarios, e.g., cross-channel \cite{vox2}, cross-lingual \cite{sdsv20,sdsv21_plan}, and far-field setting \cite{ffsvc}. In this paper, we further improve the performance of the speaker verification system with two strategies, i.e., improving the network structures with a new attention module and data cleaning for a potentially noisy dataset.

One important improvement of neural network structure is the application of the attention mechanism. Under the branch of convolutional neural networks (CNN), the squeeze-and-excitation (SE) module \cite{senet} employs the channel-wise attention to capture the task-relevant features. Convolutional block attention module (CBAM) \cite{cbam} extends the attention to the spatial dimension. CBAM sequentially infers 1-dimensional (1D) and 2-dimensional (2D) attention weights for the channel and spatial dimensions. Since the spectrogram of a speech signal is a time series, the 2D weights for spatial dimensions may not extract enough temporal information. Recently, SimAM \cite{SimAM} proposes to find the importance of each neuron by optimizing an energy function without adding extra modal parameters. The SimAM module generates 3D attentions weights for the feature map in a layer of CNN, which are more suitable for speech-related tasks. This paper uses the SimAM module in the deep speaker verification framework to achieve better performance.

Supervised learning methods usually require data with accurate annotations. Data with the noisy label may over parameterize a deep neural network (DNN) and lead to performance degradation due to the memorization effect of the DNN. However, the problem of data mislabeling is inevitable in the real-world scenario, and re-labeling can be time-consuming. To this end, we propose a simple method to iteratively filter out the noisy label and improve the performance with noisy training data. Specifically, we extract the speaker embeddings of all utterances in the same speaker. Cosine similarities of each training utterance are calculated with other segment average embeddings of the corresponding speaker. Our proposed noisy label detection method filters out audios with average cosine similarities below the predefined threshold.

To sum up, our main contributions are:

\begin{itemize}
	\item We introduce a 3-D attention module that designs an energy function to compute the weight for the ASV system. This plug-and-play module achieves the state-of-the-art (SOTA) results in the VoxCeleb test set.
	\item We also propose an iterative noisy label detection method to filter out data with unreliable labels. Compared to the strong baseline systems, this method has an additional 7\% relatively improvement.   
\end{itemize}


\section{Attention modules}
\label{sec:simam}
In this section, we will introduce the attention modules that have been successfully used in ASV and the SimAM module.

\subsection{Related works}
\subsubsection{Channel-wise squeeze-excitation}
The SE module \cite{senet} has achieved a great success in both computer vision and speech processing fields. The standard SE module uses two fully connected layers to learn the importance of different channels by first compressing and then expanding the full average channel vector to obtain channel-level weights. Given the output feature map $\mathbf{x} \in \mathbb{R}^{C \times F \times T}$ of the convolutional layer, the SE module first calculate the channel-wise mean statistics $\mathbf{e} \in \mathbb{R}^C$. The \textit{c}-th element of $\mathbf{e}$ is 
\begin{equation}
	\mathbf{e}_c=\frac{1}{F\times T}\sum_{i=1}^{F}\sum_{j=1}^{T}\mathbf{x}_{c,i,j}
\end{equation}
where $C$, $F$ and $T$ represent the channel, frequency and time dimension. The SE module then scaled this channel-wise mean by two fully connected layers to obtain the attention weights $\mathbf{s}$ of different channels:
\begin{equation}
\label{eq:s}
	\mathbf{s}=\sigma(\mathbf{W}_{2}f(\mathbf{W}_{1}\mathbf{e}+\mathbf{b}_{1})+\mathbf{b}_2),
\end{equation}
where $\mathbf{W}$ and $\mathbf{b}$ indicate the weight and bias of a linear layer, $f(\cdot)$ is the activate function of rectified linear unit (ReLU) and $\sigma(\cdot)$ is the sigmoid function. 
 
\subsubsection{Frequency-wise squeeze-excitation}
To tailor the SE module for speech processing tasks, Thienpondt \textit{et al.} \cite{idlab_sdsv20} propose the frequency-wise squeeze-excitation (fwSE) module, which aggregates global frequency information as attention weights for all feature maps. The \textit{f}-th element of of the frequency-wise mean statistics $\mathbf{e} \in \mathbb{R}^F$ is calculated as
\begin{equation}
	\mathbf{e}_{f}=\frac{1}{C\times T}\sum_{i=1}^{C}\sum_{j=1}^{T}\mathbf{x}_{i,f,j}
\end{equation}
The generation of the attention weights of the fwSE module is same as equation (\ref{eq:s}) in SE module.

\subsubsection{Convolutional block attention modules}

%
%
%
%
%
The CBAM proposed in \cite{cbam} adopts the channel attention and spatial attention submodules on the input maps of the ResNet block and has been used in the ASV task \cite{ft-cbam}. The frequency and temporal convolutional attention module (ft-CBAM) obtains the statistical vectors by extracting the average pooling and maximum pooling on frequency and time domains. The statistical vectors are mapped through a fully connected layer and then passed through the Sigmoid activation function to obtain frequency and temporal attention weights. 

%
%
%
%

%
%
%
%
%
%
%

\subsection{Simple attention module}

Based on the phenomenon of spatial suppression \cite{neuro_theory} in neuroscience\footnote{In neuroscience, the phenomenon of spatial suppression is the suppression of surrounding neurons' activities from an active neuron.}, the following energy function is defined for each neuron in a feature map $\mathbf{x} \in \mathbb{R}^{C\times H \times W}$ of a CNN layer \cite{SimAM}:
\begin{equation}
	\label{eq:energy}
	e_{t}(w_{t},b_{t},\mathbf{y}, x_{i})=(y_{t}-\hat{t})^{2}+\frac{1}{M-1}\sum^{M-1}_{i=1}(y_{o}-\hat{x}_{i})^{2}
\end{equation}
Here $\hat{t}=w_{t}t+b_{t}$ and $\hat{x}_{i}=w_{t}x_{i}+b_{t}$ are linear transforms of the target neuron $t$ and other neurons $x_{i}$ in a single channel of the feature map $\mathbf{x}$. $i$ is index over the time-frequency dimension. $y_{t}$ and $y_{o}$ are two different values for target neuron $t$ and other neurons $x_{i}$. The energy function is minimized when $\hat{t}$ equals $y_{o}$ and $\hat{x}_i$ equals $y_{t}$. Without loss of generality, the energy function in equation (\ref{eq:energy}) is simplified with binary values of $y_{t}=1$ and $y_{o}=-1$,
\begin{equation}
\begin{aligned}
	\label{eq:energy_simp}
	e_{t}(w_{t},b_{t},\mathbf{y},x_{i})=&\frac{1}{M-1}\sum^{M-1}_{i=1}(-1-(w_{t}x_{i}+b_{t}))^{2} \\
	& +(1-(w_{t}t+b_{t}))^{2}+\lambda w_{t}^{2}
\end{aligned}
\end{equation}
The above function is computationally complex in the optimization process. Luckily, equation (\ref{eq:energy_simp}) has a closed-form solution which can be obtained by differentiating $w_{t}$ and $b_{t}$. Putting $w_{t}$ and $b_{t}$ back into the energy function gives the minimal energy:
\begin{equation}
	e_{t}^{*}=\frac{4(\hat{\sigma}^{2}+\lambda)}{(t-\hat{\mu})^{2}+2\hat{\sigma}^{2}+2\lambda}
\end{equation}
where $\hat{\mu}=\frac{1}{M}\sum^{M}_{i=1}x_{i}$ and $\hat{\sigma}^{2}=\frac{1}{M}\sum^{M}_{i=1}(x_{i}-\hat{\mu})^{2}$.

Based phenomenon of spatial suppression, the lower energy of $e_{t}^{*}$ indicates the more important of the neuron $t$. The final outputs are thus obtained as:
\begin{equation}
	\widetilde{\mathbf{x}}=\sigma \left( \frac{1}{\mathbf{E}} \right) \otimes \mathbf{x},
\end{equation}
where the $\otimes$ denotes the element-wise multiplication, $\mathbf{E}$ contains all energy values of $e_{t}^{*}$ across the whole feature map, and $\sigma(\cdot)$ is the sigmoid function.

By optimizing an energy function for each neuron of the 3D feature map, the simple attention module calculates 3D attention weights without introducing extra parameters for model training. 

\section{Iterative noisy label detection}

The recent success of ASV depends on the availability of large-scale and carefully labeled supervised training data. However, the automated labeling processes may introduce noisy labels, which could degrade the system performance. Figure \ref{fig:error_spk} shows an example with noisy labels in the VoxCeleb2 development set \cite{vox2}. Different face identities are associated with the utterances of the same speaker, which indicates that these utterances are mislabeled at a high probability.

In this paper, we propose a noisy label detection approach to iteratively filter out data with noisy label in VoxCeleb dataset \cite{vox1, vox2}. The details of the proposed method is as follows:
\begin{itemize}
	\item Step 1. Given the the original training data $\mathcal{D} = \{\x_{s,v,u}\}$, where $s, v, u$ represent the indexes of speaker, video segment, and utterance respectively, let the current training data $\hat{\mathcal{D}}=\mathcal{D}$.
	\item Step 2. Train a speaker embedding network with $\hat{\mathcal{D}}$.
	\item Step 3. Extract speaker embeddings $\{\mathbf{f}_{s,v,u}\}$ for $\hat{\mathcal{D}}$. For a target embeddings $\mathbf{f}_{s,v,u}$, average the speaker embeddings with different video indexes within the same speaker:
	\begin{equation}
	\mathbf{f}_{s, \setminus v} = \frac{1}{\sum_{i \neq v} \sum_u 1} \sum_{i \neq v} \sum_u \mathbf{f}_{s,i,u}
    \end{equation}
	Calculate cosine similarities for the whole dataset $\hat{\mathcal{D}}$ as $\mathrm{cosine}(\mathbf{f}_{s,v,u}, \mathbf{f}_{s, \setminus v})$.
	\item Step 4. Generate new training data $\hat{D}$ by rejecting data samples with an average cosine similarity score that is below a predefined threshold.
	\item Step 5. Repeat step 2 to step 4 with several rounds until little utterances are below the threshold.
\end{itemize}

The final noisy label list in our experiment has been released online\footnote{Available at https://github.com/qinxiaoyi/Simple-Attention-Module-based-Speaker-Verification-with-Iterative-Noisy-Label-Detection}.

\begin{table*}[tp] \small
  \caption{The performance of different speaker verification systems. SN indicates Score normalization.}

  \label{tab:vox_result}
  \centering
  \begin{tabular}[c]{llccccccc}
    \toprule
     \multirow{2}*{\textbf{Front-end}} & \multirow{2}*{\textbf{Pooling}} & \multirow{2}*{\textbf{SN}} & \multicolumn{2}{c}{\textbf{VoxCeleb1-O}} &  \multicolumn{2}{c}{\textbf{VoxCeleb1-E}} & \multicolumn{2}{c}{\textbf{VoxCeleb1-H}} \\
     
     \cmidrule(lr){4-5} \cmidrule(lr){6-7} \cmidrule(lr){8-9} 
     
      &  & & \textbf{EER[\%]}  & \textbf{mDCF}$_{0.01}$ & \textbf{EER[\%]}  & \textbf{mDCF}$_{0.01}$ & \textbf{EER[\%]}  & \textbf{mDCF}$_{0.01}$ \\
 
   \midrule
	 ResNet34 & GSP & AS Norm & 0.851 & 0.079 & 1.054 & 0.114 & 1.825 & 0.172   \\
	 SE-ResNet34 & ASP & AS Norm & 0.776 & 0.088 & 0.921 & 0.105 & 1.703 & 0.166    \\
	 fwSE-ResNet34\cite{idlab_sdsv20} & ASP & ASNorm & 0.70 & 0.0856 & - & - & - & -    \\
	 
	 ECAPA-TDNN(C=1024) & ASP & AS Norm & 0.734 & 0.088 & 0.968 & 0.109 & 1.848 & 0.179    \\
	\midrule
	 SimAM-ResNet34 & GSP & - & 0.798 & 0.085 & 1.002 & 0.113 & 1.798 & 0.179    \\
	 SimAM-ResNet34 & GSP & AS Norm& 0.718 & 0.071 & 0.993 & 0.103 & 1.647 & 0.159    \\
     SimAM-ResNet34 & ASP & - & 0.729 & 0.095 & 0.959 & 0.104 & 1.782 & 0.183  \\	
	 SimAM-ResNet34 & ASP & AS Norm & \bf{0.675} & \bf{0.077} & \bf{0.867} & \bf{0.094} & \bf{1.567} & \bf{0.155}    \\
	 \quad +INLD (2 rounds) & ASP& - & 0.670 & 0.082 & 0.914 & 0.099 & 1.638 & 0.163 \\
	 \quad +INLD (2 rounds) & ASP& AS Norm & \bf{0.643} & \bf{0.067} & \bf{0.842} & \bf{0.089}& \bf{1.491} & \bf{0.146} \\
     \bottomrule
     \end{tabular}
\end{table*}

\section{Experiments}

\subsection{Experimental setting}

\subsubsection{Dataset}
Speaker embedding models are trained on the development set of VoxCeleb 2 \cite{vox2} that consists of 5,994 speakers with 1,092,009 utterances. Evaluation is performed on the VoxCeleb 1 dataset \cite{vox1}. We report the speaker verification results on three trial lists as defined in \cite{vox2}: (1) VoxCeleb 1-O: original trial list containing 37,611 trials from 40 speakers; (2) Voxceleb 1-E: extended trial list containing 579,818 trials from 1251 speakers; (3) Voxceleb 1-H: hard trial list containing 550,894 trials from 1190 speakers.

\subsubsection{Data augmentation}
We adopt the on-the-fly data augmentation \cite{cai_on-the-fly} to add additive background noise or convolutional reverberation noise for the time-domain waveform. The MUSAN \cite{musan} and RIR Noise \cite{RIR} datasets are used as noise sources and room impulse response functions, respectively. To further diversify training samples, we apply amplification or playback speed change (pitch remains untouched) to audio signals.
Also, we apply speaker augmentation with speed perturbation \cite{speed_perturb_spk,sdsv21_qin,dku_voxsrc20}. Specifically, we speed up or down each utterance by a factor of 0.9 or 1.1, yielding shifted pitch utterances that are considered from new speakers. As a result, the training data includes 3,276,027 (1,092,009$\times$3) utterances from 17,982 (5,994$\times$3) speakers. 

\subsubsection{Model training and evaluation}

For feature extraction, logarithmical Mel-spectrogram is extracted by applying 80 Mel filters on the spectrogram computed over Hamming windows of 20ms shifted by 10ms.

We adopt the SOTA ASV models, namely ResNet34, SE-ResNet34 and ECAPA-TDNN, as the baselines. The implementation of ResNet34 is the same as in \cite{cai_exploring_2018}. SE-ResNet34 adds the SE module to ResNet34. For ECAPA-TDNN \cite{ecapatdnn}, 1024 feature channels are used to scale up the network and the dimension of the bottleneck in the SE-Block is set to 256.
The encoding layer is based on global statistic pooling (GSP) or attentive statistics pooling (ASP) \cite{asp_pooling}. The speaker embedding is with a dimension of 256.
Additive angular margin (AAM) loss \cite{arcface} with re-scaling factor $s$ of 32 and angular margin $m$ of 0.2 is used to train all systems.
The detail of other training strategy, hyperparameters and models configuration follows \cite{dku_voxsrc20,sdsv21_qin}.


 During evaluation, cosine similarity is used as the scoring function. All scores are normalized with adaptive symmetric score normalization (ASNorm) \cite{asnorm}. The size of the imposter cohort is set to 400. 

\begin{table}[tp] \small
  \caption{Model Size.}
  \label{tab:model_size_3d}
  \centering
  
  \begin{tabular}[c]{lcc}
    \toprule
    \textbf{Model} & Parameters (M) \\
	\midrule
	ECAPA\_TDNN & 20.12 \\
	ResNet34 GSP & 21.54  \\
	SimAM-ResNet34 GSP & \bf{21.54}  \\
	SE-ResNet34 ASP  & 25.53  \\
	SimAM-ResNet34 ASP & \bf{25.21}  \\
	\bottomrule
     \end{tabular}
\end{table}

\begin{table}[tp] \small
  \caption{The performance of iterative noisy label detection.}
  \label{tab:noisy_result}
  \centering
  
  \begin{tabular}[c]{@{\ \ \ }l@{\ \ \ }c@{\ \ \ }c@{\ \ \ }c@{\ \ \ }c@{\ \ \ }}
    \toprule
    \textbf{Model} & \textbf{ Noisy Utt} & \textbf{Threshold} & \textbf{EER[\%]} & \textbf{mDCF$_{0.01}$}\\
	
	\midrule
	Initial round & 0 & - & 0.7286 & 0.095  \\
	Round 1 & 17697 & 0.4 & 0.6860 & 0.083  \\
	Round 2 & 10646 & 0.5 & \bf{0.6701} & \bf{0.082}  \\
	\bottomrule
	
     \end{tabular}
\end{table}

\subsection{Experimental results}
Verification performances are measured by EER and the minimum normalized detection cost function (mDCF) with $P_\mathrm{target} = 10^{-2}$ and $C_\mathrm{FA} = C_\mathrm{Miss} = 1$.

Table \ref{tab:vox_result} presents the verification results. Integrating either SE or SimAM into ResNet can significantly boost the performance. SimAM-ResNet34 obtains a 5\% relative improvement on top of SE-ResNet34 without adding extra parameters. The SimAM-ResNet34 has achieved 0.675\% EER on the VoxCeleb1 original test set as a single system. Table \ref{tab:model_size_3d} shows a comparison of model size.

Table \ref{tab:noisy_result} shows the results of SimAM-ResNet34 after two rounds of iterative training and label refinement. The first round of iterative noisy label detection rejects 17,697 utterances with the detection threshold of 0.4. Although there are few utterances with cosine similarity below 0.4 after the first round, we observe some noisy utterances with scores ranged from 0.4 to 0.5. Thus, we increase the threshold to 0.5 and further exclude 10,646 unreliable utterances. After two rounds of noisy label detection, EER improves from 0.73\% to 0.67\% compared with the initial model.

\section{Conclusions}

In this paper, we introduce the simple attention module to speaker verification. SimAM calculates 3D attention weights without introducing extra modal parameters. Experiments on VoxCeleb 1 test set show that SimAM obtains 5\% relative EER reduction compared to the baseline model. In addition, to handle the noisy label, we propose an iterative noisy label detection approach to refine the training data labels. The proposed noisy label detection method achieves another 7\% relative EER reduction. 

\small
\bibliographystyle{IEEEbib}
\bibliography{strings,refs}

\begin{thebibliography}{10}

\bibitem{xvector}
D.~Snyder, D.~Garcia-Romero, G.~Sell, D.~Povey, and S.~Khudanpur,
\newblock ``x-{vectors}: {Robust} {DNN} {Embeddings} for {Speaker}
  {Recognition},''
\newblock in {\em Proc. {ICASSP}}, 2018, pp. 5329--5333.

\bibitem{ftdnn}
D.~Povey, G.~Cheng, Y.~Wang, K.~Li, H.~Xu, M~Yarmohammadi, and S.~Khudanpur,
\newblock ``Semi-orthogonal low-rank matrix factorization for deep neural
  networks,''
\newblock in {\em Proc. Interspeech 2018}, 2018, pp. 3743--3747.

\bibitem{resnet}
K.~He, X.~Zhang, S.~Ren, and J.~Sun,
\newblock ``{Deep Residual Learning for Image Recognition},''
\newblock in {\em Proc. CVPR}, 2016, pp. 770--778.

\bibitem{res2net}
S.~Gao, M.~Cheng, K.~Zhao, X.~Zhang, M.~Yang, and P.~Torr,
\newblock ``Res2net: A new multi-scale backbone architecture,''
\newblock {\em arXiv:1904.01169}, 2019.

\bibitem{vox2}
J.S. Chung, A.~Nagrani, and A.~Zisserman,
\newblock ``Voxceleb2: {Deep} {Speaker} {Recognition},''
\newblock in {\em Proc. Interspeech}, 2018.

\bibitem{sdsv20}
H.~Zeinali, K.A. Lee, J.~Alam, and L.~Burget,
\newblock ``{SdSV Challenge 2020: Large-Scale Evaluation of Short-Duration
  Speaker Verification},''
\newblock in {\em Proc. Interspeech}, 2020, pp. 731--735.

\bibitem{sdsv21_plan}
H.~Zeinali, K.A. Lee, J.~Alam, and L.~Burget,
\newblock ``{Short-duration Speaker Verification (SdSV) Challenge 2021: the
  Challenge Evaluation Plan},''
\newblock {\em arXiv:1912.06311}.

\bibitem{ffsvc}
X.~Qin, M.~Li, H.~Bu, W.~Rao, K.R. Das, S.~Narayanan, and H.~Li,
\newblock ``{The INTERSPEECH 2020 Far-Field Speaker Verification Challenge},''
\newblock in {\em Proc. Interspeech}, 2020, pp. 3456--3460.

\bibitem{senet}
J.~Hu, L.~Shen, and G.~Sun,
\newblock ``{Squeeze-and-Excitation Networks},''
\newblock in {\em Proc. CVPR}, 2018.

\bibitem{cbam}
S.~Woo, J.~Park, Y.~Lee, and I.S. Kweon,
\newblock ``{CBAM:} convolutional block attention module,''
\newblock {\em arxiv 1807.06521}, 2018.

\bibitem{SimAM}
L.~Yang, R.~Zhang, L.~Li, and X.~Xie,
\newblock ``Simam: A simple, parameter-free attention module for convolutional
  neural networks,''
\newblock in {\em Proc. ICML}, 2021, pp. 11863--11874.

\bibitem{idlab_sdsv20}
J.~Thienpondt, B.~Desplanques, and K.~Demuynck,
\newblock ``{Integrating Frequency Translational Invariance in TDNNs and
  Frequency Positional Information in 2D ResNets to Enhance Speaker
  Verification},''
\newblock in {\em Proc. Interspeech}, 2021, pp. 2302--2306.

\bibitem{ft-cbam}
S.~Yadav and A.~Rai,
\newblock ``Frequency and temporal convolutional attention for text-independent
  speaker recognition,''
\newblock in {\em Proc. {ICASSP}}, 2020, pp. 6794--6798.

\bibitem{neuro_theory}
B.S. Webb, N.T. Dhruv, S.G. Solomon, C.~Tailby, and P.~Lennie,
\newblock ``Early and late mechanisms of surround suppression in striate cortex
  of macaque,''
\newblock {\em The Journal of neuroscience}, vol. 25, no. 50, pp. 11666--11675,
  2005.

\bibitem{vox1}
A.~Nagrani, J.S. Chung, and A.~Zisserman,
\newblock ``Voxceleb: {A} {Large}-{Scale} {Speaker} {Identification}
  {Dataset},''
\newblock in {\em Proc. Interspeech}, 2017, pp. 2616--2620.

\bibitem{cai_on-the-fly}
W.~{Cai}, J.~{Chen}, J.~{Zhang}, and M.~{Li},
\newblock ``{On-the-Fly Data Loader and Utterance-Level Aggregation for Speaker
  and Language Recognition},''
\newblock {\em IEEE/ACM Transactions on Audio, Speech, and Language
  Processing}, vol. 28, pp. 1038--1051, 2020.

\bibitem{musan}
D.~Snyder, G.~Chen, and D.~Povey,
\newblock ``{MUSAN}: {A} {Music}, {Speech}, and {Noise} {Corpus},''
\newblock {\em arXiv:1510.08484}.

\bibitem{RIR}
T.~Ko, V.~Peddinti, D.~Povey, M.L. Seltzer, and S.~Khudanpur,
\newblock ``A study on data augmentation of reverberant speech for robust
  speech recognition,''
\newblock in {\em Proc. ICASSP}, 2017, pp. 5220--5224.

\bibitem{speed_perturb_spk}
H.~Yamamoto, Lee. K.A., K.~Okabe, and T.~Koshinaka,
\newblock ``{Speaker Augmentation and Bandwidth Extension for Deep Speaker
  Embedding},''
\newblock in {\em Proc. Interspeech}, 2019, pp. 406--410.

\bibitem{sdsv21_qin}
X.~Qin, C.~Wang, Y.~Ma, M.~Liu, S.~Zhang, and M.~Li,
\newblock ``{Our Learned Lessons from Cross-Lingual Speaker Verification: The
  CRMI-DKU System Description for the Short-Duration Speaker Verification
  Challenge 2021},''
\newblock in {\em Proc. Interspeech}, 2021, pp. 2317--2321.

\bibitem{dku_voxsrc20}
W.~Wang, D.~Cai, X.~Qin, and M.~Li,
\newblock ``{The DKU-DukeECE Systems for VoxCeleb Speaker Recognition Challenge
  2020},''
\newblock {\em arXiv.2010.12731}.

\bibitem{cai_exploring_2018}
W.~Cai, J.~Chen, and M.~Li,
\newblock ``Exploring the {Encoding} {Layer} and {Loss} {Function} in
  {End}-to-{End} {Speaker} and {Language} {Recognition} {System},''
\newblock in {\em Proc. Speaker Odyssey}, 2018, pp. 74--81.

\bibitem{ecapatdnn}
D.~Desplanques, J.~Thienpondt, and K.~Demuynck,
\newblock ``{ECAPA-TDNN: Emphasized Channel Attention, Propagation and
  Aggregation in TDNN Based Speaker Verification},''
\newblock in {\em Proc. Interspeech}, 2020, pp. 3830--3834.

\bibitem{asp_pooling}
K.~Okabe, T.~Koshinaka, and K.~Shinoda,
\newblock ``{Attentive Statistics Pooling for Deep Speaker Embedding},''
\newblock {\em Proc. Interspeech}, 2018.

\bibitem{arcface}
J.~{Deng}, J.~{Guo}, N.~{Xue}, and S.~{Zafeiriou},
\newblock ``{ArcFace: Additive Angular Margin Loss for Deep Face
  Recognition},''
\newblock in {\em Proc. CVPR}, 2019, pp. 4685--4694.

\bibitem{asnorm}
P.~Matějka, O.~Novotný, O.~Plchot, L.~Burget, M.D. Sánchez, and
  J.~Černocký,
\newblock ``Analysis of {Score} {Normalization} in {Multilingual} {Speaker}
  {Recognition},''
\newblock in {\em Proc. Interspeech}, 2017.

\end{thebibliography}

\end{document}